# Triple Helix synergy and patent dynamics. Cross country comparison

Inga Ivanova[1] and Grzegorz Rzadkowski[2]

**Abstract**

We use a computationally efficient technique of logistic Continuous Wavelet Transform (CWT) to analyze patent data for Switzerland, Germany, USA, and Brazil for the period 1980-2021. We found that patent growth dynamics follows the dynamics of innovation system synergy in the framework of the Triple Helix model of innovations where observed non-linear actors' interactions are provided by biased information exchange between heterogeneous actors. Suggested approach reveals the latent trend structure in patent and innovation dynamics and may help policymakers identify the potential drivers of patent and innovation activity and form informed policy for boosting innovation development. The paper also provides a foundation for future research in different fields studying complex systems of interacting heterogeneous agents.

## 1 Introduction

Science, technology and innovations are considered the major drivers of sustainable economic development of countries in post-industrial economy. Economic growth increasingly relies on innovation activity (OECD, 2018). Continuous innovations are generated via collaborative and interactive relationship of economic agents, policy makers, science research organizations, and other innovation system participants. Innovation system can be considered as complex adaptive system, is subject to non-linear mechanisms which govern system evolution (e.g. Dosi 2014, Russell and Smorodinskaya, 2018).

---
[1] Institute for Statistical Studies and Economics of Knowledge, HSE University, 20 Myasnitskaya St., Moscow, 101000, Russia; inga.iva@mail.ru ; ORCID: 0000-0002-5441-5231.
[2] Department of Finance and Risk Management, Warsaw University of Technology, Narbutta 85, 02-524 Warsaw, Poland e-mail: grzegorz.rzadkowski@pw.edu.pl

The most important property of innovation system is its effectiveness. Estimation of the effectiveness of innovation system is of primary concern to researches and policy makers. Effectiveness can be measured quantitatively with respect to specific representation of innovation system, such as, e.g., Triple Helix (TH) model of innovations (Etzkowitz and Leydesdorff, 1995). Neo-institutional (actors oriented) TH model explains the phenomenon of creating and introducing innovations via interaction of three major actors: University, Industry, and Government. Neo-institutional model can be complemented by neo-evolutional model with accent on corresponding actors' carrying functions – novelty production, wealth generation, and normative control – which provide three different sub-dynamics (Leydesdorff, 2000; Leydesdorff & Zawdie, 2010). A system with three sub-dynamics can endogenously generates complex non-linear behavior (Ivanova and Leydesdorff, 2014a). These sub-dynamics driven by analytically different mechanisms can be considered as selection environments representing mechanisms of social coordination which can interact synergetically in shaping innovation system Leydesdorff, L. (2010). When two coordination mechanisms interact they can shape each other in co-evolution and be "locked-in" in historical trajectory. Three bi-lateral trajectories can shape three-lateral regime.

In the framework of TH model, the effectiveness of innovation system can be measured quantitatively in terms of probabilistic entropy calculated as mutual information in three dimensions. This entropy can be either positive or negative. Positive entropy indicates increase of uncertainty that prevails at the system level, as a result of historical evolution, and negative entropy indicates reduction of uncertainty. This negative entropy can be considered as a measure of synergy of TH actors' interaction generated within the system (e.g. Park, Leydesdorff, 2010; Leydesdorff, Park, Lenguel, 2014). Synergy implies that total system's output obtained from actors' interaction exceeds a sum of partial outputs from non-interacting actors and indicates a potential for innovation activity that fosters the progress in all the spheres of novelty production, wealth generation, and normative control, which can be considered crucial for the strength of an innovation system.

Synergy is related to a structural level of innovation system and is not an indicator of knowledge generation or economic output. However, as a structural measure, it determines the overall performance potential of the system. One can assume that synergy, as an information measure used for an integral assessment of the quality of an innovation system, will also influence other indicators characterizing its effectiveness. E.g. TH synergy correlates with the economic complexity index, calculated on the basis of countries' exported products, and used to measure countries' relative economic complexity and predict future economic growth (Hidalgo and Hausmann, 2009, Ivanova, 2022a).

TH metaphor, as an example of inherently non-linear system, apart from innovation studies, is also applicable to other fields, such as the COVID-19 pandemic spread, financial markets, and rumors propagation (Ivanova, 2024).

Innovation performance can be estimated with respect to different spheres. E.g. the Global Innovation Index comprises about 80 indicators distributed to different clusters (WIPO, 2023). In the present paper we take patents as units of innovation performance analysis. Patents play an important role in the process of innovation and technological change being a significant factor of economic development (e.g. Pavitt, 1985; Dang & Motohashi, 2015). Patents are framed in different contexts. Being an output of knowledge production they can be considered as indicators of inventions (Griliches, 1990) which shouldn't be mixed with innovations, because not all patents are implemented in innovative products and not all innovations (e.g. organizational innovations) can be patented, but serve as input to the process of innovation as they "*represent one important outcome of companies' efforts at innovation*" (Jung *et al*., 2008, at. p.21). Also enhancements of intellectual property legislative sphere are retained in institutional framework and lead to an increase in the number of patents (Lerner, 2002). Thus all the three domains of TH actors' activities – novelty production, wealth generation, and normative control – contribute to patent production growth.

This paper introduces a way of assessing countries' patent performance in terms of TH synergy. Synergy is a comprehensive measure of innovation system performance. One can expect that observed patents synergy cycles also relate to cycles in other innovation related domains. Moreover synergy analysis can provide the footprints of the trade-offs between historical organization and self-organization. Historical organization manifests itself as technological trajectories are organized locally in "landscapes." Self-organization implies that selection environments self-organize into regimes in terms of different codes of communication entertained by different innovation system actors. So that innovation cycles are driven by these two tendencies (Leydesdorff, 2021). Our hypothesis is that there should be close relation between synergy generated within the innovation system and patent production. More specifically patent dynamics should follow the patterns predicted by synergy dynamics. To test the hypothesis we develop a technique for patent trends analysis comprising continuous logistic wavelet transformation (CWT). Rzadkowski and Figlia (2021) exploited the second-order logistic wavelets for modelling the spread of COVID 19 pandemic in different countries. The parameters of the appropriate logistic functions were obtained by applying the continuous wavelet transform to the second differences of the total number of COVID 19 cases. Similar method we will use also in the present paper.

To the best of our knowledge longitudinal patent data had been never examined via wavelet decomposition. Logistic functions derivatives correspond to soliton solutions of Korteweg-de Vries (KdV) equation which naturally appears in TH model (Ivanova (2022b, 2022c). Decomposition of patent data into a sum of logistic functions can give information about mechanisms that drive innovation system dynamics. The balance between historical entropy generation and the knowledge-based generation of options can be measured in terms of positive and negative contributions to the prevailing uncertainty. CWT transform allows distinctly differentiate between these two tendencies.

Our results suggest that patent dynamics is closely correlated with TH synergy dynamics, so that patent trends evolve in the way predicted by synergy evolution. This evolution in turn is provided by information exchange among heterogeneous actors which comprise TH innovation system. Here the term "heterogeneous" implies that different actors supply different meanings to the same information due to different meaning processing structures.

The paper findings and methodology in a narrow sense can lead to a better understanding of patent, innovation, and innovation systems dynamics and help policymakers to undertake measures to accelerate economic growth and competitiveness. In a wider sense the applicability of proposed framework is not limited to patent or innovation studies. It can be used in much more fields of applications to study the dynamics of different complex systems where systems in question are presented as a set of interacting heterogeneous actors. We expect to develop this approach in future studies.

The rest of the paper is structured as follows: In section 2 we shortly explain the underlying theory. In section 3 we specify in details the method of CWT used for developing the patent data in logistic wavelets. Section 4 describes the obtained results which are further discussed in section 5 alongside with the main findings and conclusions.

## 2      Theory

In this section we briefly describe the underlying theory of synergy generation in TH configuration. Our hypothesis is that the amount of patents generated within the national innovation system (NIS) is proportional to the synergy value of system's actors interaction.. Synergy in the TH configuration can be measured quantitatively. Leydesdorff (2003) suggested use mutual information in three (or more) dimensions, defined as overlapping uncertainties in three variables, as an indicator of synergy of actors' interaction. For the case of two distributions (Fig.2.1)

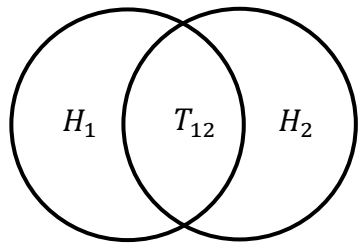

**Figure 2.1**: Overlapping uncertainties in two variables

$$H_{12} = H_1 + H_2 - T_{12} \tag{1}$$

The value $T_{12}$ is Shannon's mutual information. When one adds third dimension (Fig.2.2)

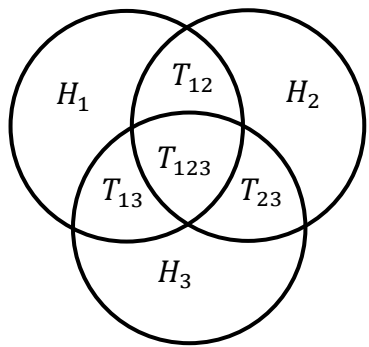

**Figure 2.2**: Overlapping uncertainties in three variables

the formula for mutual uncertainty is changed as:

$$H_{123} = H_1 + H_2 + H_3 - H_{12} - H_{13} - H_{23} - T_{123} \tag{2}$$

Here: $H_i = -\sum p_i \log(p_i)$, $H_{ij} = -\sum p_{ij} \log(p_{ij})$, $H_{ijk} = -\sum p_{ijk} \log(p_{ijk})$; $p_i$, $p_{ij}$, $p_{ijk}$ – corresponding probabilities are defined as $p_i = \frac{n_i}{N}$; $p_{ij} = \frac{n_{ij}}{N}$; $p_i = \frac{n_{ijk}}{N}$; $N$ – total number of items, $n_{ijk}$ – number of indexed items in distributions and overlaps. The sign of lastly added term alternates with each added dimension (Yeung, 2008). The sign change can be avoided by introducing "excess" information value (Leydesdorff and Ivanova, 2014) which is defined as following:

$$Y_1 = H_1$$
$$Y_2 = H_2$$
$$Y_{12} = H_1 + H_1 + T_{12} = H_{12} + 2T_{12} \qquad (3)$$

Here the value $T_{12}$ is added to the sum, which is contrary to formula (1). One can further define the value

$$R_{12} = H_1 + H_2 - Y_{12} \qquad (4)$$

and it follows that

$$R_{12} = -T_{12}$$

One can also show that for higher dimensions the sign between $R$ and $T$ sequentially changes

$$R_{123} = T_{123}$$

$$R_{1234} = -T_{1234}$$

$$\dots$$

which compensate the sign alteration of mutual information $T$ in higher dimensions, so that the problem of sign change is eliminated. Since $T_{12}$ is Shannon's mutual information which is always positive $R_{12}$ is negative. While $T_{12}$ diminishes resulting uncertainty by the amount of overlap, $R_{12}$ augments resulting uncertainty by the same amount. Therefore, $R_{12}$ can be considered as excess uncertainty or redundancy. Negative redundancy indicates the reduction of total uncertainty, considered as synergy. The mechanism of "excess" overlapping can be attributed to generation of additional options in the overlaps due to different ways of processing the information by different TH actors, so that the same information can be supplied with different meanings. Meanings, in turn, generate additional options which can be realized in future system evolution. Detailed description of the theory can be found in Leydesdorff (2021). Redundancy $R$ can be either positive or negative because it is a result of interaction between two dynamics: historical organization and evolutionary self-organization which can be defined by the following formula as (Ivanova, Leydesdorff, 2014b):

$$R(t) = P^2(t) - Q^2(t) \qquad (5)$$

First (positive) term in Eq. (2.2) accounts for historical organization and adds to the uncertainty, second (negative) term is due to evolutionary self-organization and leads to the reduction of uncertainty.

Redundancy changes over time along the trajectory which can be considered an average to smaller fluctuations which it contains. When redundancy oscillates around the average probabilities included in the formula for information entropy are solutions for harmonic or non-harmonic oscillator equation (Dubois, 2019; Ivanova, 2022b). By considering anticipatory dynamics of the TH system one obtains generalized Korteweg-de-Vries (KdV) equation for corresponding probability densities: $P = \frac{dp}{dX}$:

$$P_T + 6PP_X + P_{XXX} + C_1 = 0 \qquad (6)$$

One soliton solution of Eq. (2.3) is:

$$P(X,T) = 2\left(\frac{\kappa}{2}\right)^2 ch^{-2}\left[\frac{\kappa}{2}\left(X - 4\left(\frac{\kappa}{2}\right)^2 T + 3C_1T^2\right)\right] - C_1T \qquad (7)$$

While probability evolution satisfies Equation (6), redundancy evolution is governed by similar equation (Appendix 2):

$$4R_T - 2RR_X + R_{XXX} + c_1 = 0$$

For  chain of solitons the ratio of soliton amplitudes to the time shifts is constant

$$A_i/_{T_i} = const \qquad (8)$$

In other words, the chain of solitons forms a linear trend. Corresponding dependence is traced in financial markets trends, Covid-19 infectious disease spread, rumors propagation (Ivanova, 2022a, 2022b, 2024). Here we will check whether this relation holds for patent longitudinal time series.

## 3    Methodology

We analyze cumulative patent data with help of logistic wavelet transformation. Wavelets are compact wave-like oscillations presenting a family of functions that are local in time and frequency, and in which all functions are derived from one by translating and dilating it along the time axis. The reason of using wavelet transformation is that it is more suitable for analyzing shorter time data than e.g. short Fourier transformation. Instead of a number of harmonics of Fourier transformation the data can be presented as a sum of just few waves. Also wavelet scalograms are fundamentally different from Fourier spectra in that they provide a clear reference of the spectrum of various signal features to time.

In this paper, we use the method of decomposition into logistic components, based on the analysis of the image of the continuous wavelet transform (CWT), applied to the second differences of the original time series. It allows for the estimation of all parameters of individual logistic components directly from the CWT scalogram. A scalogram is a three-dimensional graph in which the values of the dependent variable, i.e. CWT, are presented using colors. The maximum value of the CWT's modulus allows for estimating the saturation level of the logistic curve. The values of the remaining two parameters of the logistic function can be read from the scalogram axis. This method is an illustration and confirmation of the research results undertaken in the paper by Mallat and Hwang (1992), showing the possibility of finding the original function by examining properties of the modulus maxima of its CWT image.

Our method can also be treated as a complement to the method of decomposition into logistic components presented in the paper by Meyer *et al*., (1999), where the original time series and its first differences are analyzed.

### 3.1 Logistic equation

The logistic equation defining the logistic function $x = x(t)$ has the form (cf. Rzadkowski and Figlia, 2021)

$$x'(t) = \frac{s}{x_{sat}} x(x_{sat} - x), \quad x(0) = x_0 \tag{9}$$

where $t$ is time, $s > 0$-steepness (slope coefficient) and $x_{sat} \neq 0$-saturation level. We assume moreover that $x_0$ has the same sign as $x_{sat}$ and lies in the interval between 0 and $x_{sat}$.

After solving (9) we obtain the logistic function in the form

$$x(t) = \frac{x_{sat}}{1 + e^{-s(t - t_0)}} \tag{10}$$

where $t_0$ is the inflection point associated with the initial condition $x(0) = x_0 = \frac{x_{sat}}{1 + e^{st_0}}$, then

$t_0 = \frac{1}{s} log \left( \frac{x_{sat} - x_0}{x_0} \right), x(t_0) = \frac{x_{sat}}{2}.$

Equation (9) is a special case of the Riccati equation with constant coefficients

$$x'(t) = r(x - x_1)(x - x_2) \tag{11}$$

where constants $r \neq 0$, $x_1$, $x_2$ can be real or complex numbers.

If $x = x(t)$ is the solution of (11) then its $n$th derivative $x^{(n)}(t)$ $(n = 2,3,4, \dots)$ is a polynomial of the function $x(t)$ (Rzadkowski, 2006, 2008; Franssens, 2007):

$$x^{(n)}(t) = r^n \sum_{k=0}^{n-1} \binom{n}{k}(x - x_1)^{k+1}(x - x_2)^{n-k} \tag{12}$$

for $n = 2,3,\dots$, where $\binom{n}{k}$ denotes Eulerian number (the number of permutations $\{1,2,\dots,n\}$ having exactly $k$, $(k = 0,1,2,\dots,n-1)$ ascents (Graham *et al.,* 1994) .

### *3.2   Wavelets*

Let us now recall some general facts about wavelet theory (cf. Daubechies, 1992; Meyer and Ryan, 1996; Meyer, Y., 1997) which we will use later. A wavelet or mother wavelet (Daubechies, 1992, p.24) is an integrable function $\psi \in L^1(\mathbb{R})$ with the following admissibility condition:

$$C_\psi = 2\pi \int_{-\infty}^{\infty} |\xi|^{-1} \left| \hat{\psi}(\xi) \right|^2 d\xi < \infty \tag{13}$$

where $\hat{\psi}(\xi)$ is the Fourier transform of $\psi$

$$\hat{\psi}(\xi) = \frac{1}{\sqrt{2\pi}} \int_{-\infty}^{\infty} \psi(x) e^{-i\xi x} dx$$

Since the function $\psi \in L^1(\mathrm{R})$, then $\hat{\psi}(\xi)$ is a continuous function, and condition (15) is satisfied only when $\hat{\psi}(0) = 0$ or $\int_{-\infty}^{\infty} \psi(x) dx = 0$. On the other hand, Daubechies (1992, p.24) shows that condition $\int_{-\infty}^{\infty} \psi(x) dx = 0$ together with the second condition, slightly stronger than integrability, namely $\int_{-\infty}^{\infty} |\psi(x)|(1 + |x|)^\alpha dx < \infty$, for some $\alpha > 0$ are sufficient for (15). Usually much more is assumed about the function $\psi$, so from a practical point of view the conditions $\int_{-\infty}^{\infty} \psi(x) dx = 0$ and (15) are equivalent. Suppose furthermore that $\psi$ is also square integrable, $\psi \in L^2(\mathrm{R})$ with the norm

$$\|\psi\| = \left( \int_{-\infty}^{\infty} |\psi(x)|^2 dx \right)^{1/2}$$

Using the mother wavelet, by dilating and translating, a double-indexed family of wavelets is obtained

$$\psi^{a,b}(x) = \frac{1}{\sqrt{a}} \psi \left( \frac{x-b}{a} \right)$$

where $a, b \in \mathbb{R}$, $a > 0$. The normalization has been chosen so that $\|\psi^{a,b}\| = \|\psi\|$ for all $a,b$. In order to compare different wavelet families, they are usually normalized, $\|\psi\| = 1$. The Continuous Wavelet Transform (CWT) of a function $f \in L^2(\mathbb{R})$ with respect to a given wavelet family is defined as

$$(T^{wav}f)(a,b) = \langle f, \psi^{a,b} \rangle = \int_{-\infty}^{\infty} f(x)\,\psi^{a,b}(x)dx \tag{14}$$

### 3.3 Logistic wavelets

Rzadkowski and Figlia (2021) defined an unnormalized mother logistic wavelet of order n as the $n$th derivative of the logistic function $x(t) = \frac{1}{1+e^{-t}}$. Since $x'(t) = x(t)\big(1 - x(t)\big) = -x(t)(x(t) - 1)$ then from Eq. (12) we get

$$x^n(t) = (-1)^n \sum_{k=0}^{n-1} \binom{n}{k} x^{k+1}(x-1)^{n-k} = \sum_{k=0}^{n-1}(-1)^k \binom{n}{k} x^{k+1}(1-x)^{n-k} \tag{15}$$

for n=2,3, …

However, in order to compare different types of wavelets among themselves, it is more convenient to use normalized logistic wavelets.

**Lemma 1.** *For the nth derivative $x^n(t)$ of the function $x(t) = \frac{1}{1+e^{-t}}$ it holds (Appendix 1):*

$$\int_{-\infty}^{\infty} \big(x^{(n)}(t)\big)^2 dt = (-1)^{n-1} B_{2n} = |B_{2n}| \tag{16}$$

where $B_{2n}$ is the $(2n)$th Bernoulli number.

*Proof.*

Examining the soliton solutions of the Korteveg-de Vries equation, Grosset and Veselov (2005) obtained an interesting relationship between these solutions and the Bernoulli numbers

$$\int_{-\infty}^{\infty} \left(\frac{d^{n-1}}{dt^{n-1}}\frac{1}{cosh^2t}\right)^2 dt = (-1)^{n-1}2^{2n+1}B_{2n} \qquad (17)$$

Other proofs of the Grosset-Veselov formula (17) can be found in (Boyadzhiev, 2007; Rzadkowski, 2010). The formula (16) follows from the formula (17) because (we put $\tau = 2t$ at the end):

$$\int_{-\infty}^{\infty} \left(\frac{d^{n-1}}{dt^{n-1}}\frac{1}{cosh^2t}\right)^2 dt = \int_{-\infty}^{\infty} \left(\frac{d^{n-1}}{dt^{n-1}}\frac{4e^{-2t}}{(1+e^{-2t})^2}\right)^2 dt = 4\int_{-\infty}^{\infty} \left(\frac{d^n}{dt^n}\frac{1}{1+e^{-2t}}\right)^2 dt =$$

$$4(2^n)^2 \int_{-\infty}^{\infty} \left(x^{(n)}(2t)\right)^2 dt = 2(2^n)^2 \int_{-\infty}^{\infty} \left(x^{(n)}(\tau)\right)^2 d\tau \qquad (18)$$

Comparing (18) with (17) we get (16).

Bernoulli number $B_n$ vanishes for all odd numbers $n \geq 3$. The first few non-zero Bernoulli numbers are as follows $B_0 = 1$, $B_1 = -\frac{1}{2}$, $B_2 = \frac{1}{6}$, $B_4 = -\frac{1}{30}$, $B_6 = \frac{1}{42}$, $B_8 = -\frac{1}{30}$, $B_{10} = \frac{5}{66}$, $B_{12} = -\frac{691}{2730}$ (see Duren (2012), Ch. 11).

From Lemma 1 we conclude that the logistic mother wavelet $\psi_n(t)$ of order $n$ ($n = 2, 3, ...$) defined as

$$\psi_n(t) = \frac{(-1)^n}{\sqrt{|B_{2n}|}}x^{(n)}(t) \qquad (19)$$

has been normalized with the norm $\|\psi_n\| = \|\psi_n\|_{L^2} = 1$.

Since $B_4 = -1/30$, $\binom{3}{0} = 1$, $\binom{3}{1} = 4$, $\binom{3}{2} = 1$, than from equation (23) it follows that the normalized mother wavelet $\psi_2(t)$ (see Fig. 3) is of the form

$$\psi_2(t) = -\frac{\sqrt{30}}{1+e^{-t}}\left(1 - \frac{1}{1+e^{-t}}\right)\left(1 - \frac{2}{1+e^{-t}}\right) = \frac{\sqrt{30}(-e^{-3t}-e^{-t})}{(1+e^{-t})^3} \tag{20}$$

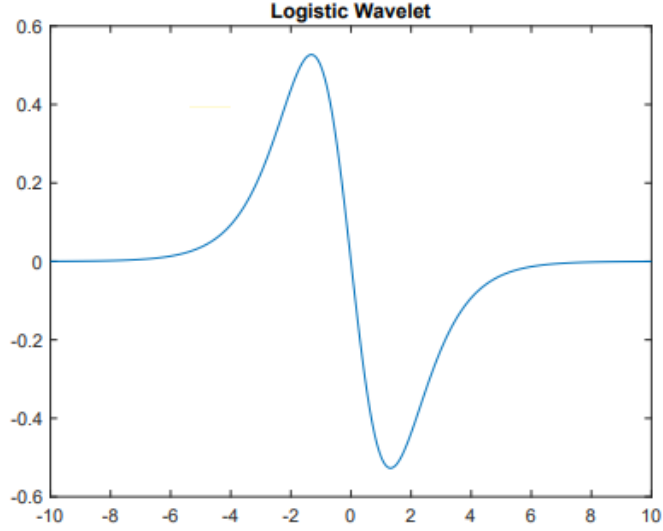

Figure 3: Wavelet $\psi_2$

We create, by dilating and translating, a doubly indexed family of wavelets (the children logistic secondorder wavelets)

$$\psi_2^{a,b}(t) = \frac{1}{\sqrt{a}}\psi_2\left(\frac{t-b}{a}\right) \tag{21}$$

where $a, b \in \mathbb{R}, a > 0$.

In the remainder of the paper we will also use functions of the form

$$f(t) = c + dt + \frac{x_{sat}}{1 + exp\left(-\frac{t-b}{a}\right)} \tag{22}$$

where a linear function has been added to the logistic function ($c$, $d$ are real constants). Taking the second derivative of the function $f(t)$ (22) we see that the linear part vanishes and by definition (20) we have

$$f''(t) = \frac{x_{sat}}{\sqrt{30}a^{\frac{3}{2}}}\psi_2^{a,b}(t)$$

**Lemma 2.** *The continuous wavelet transform (14) of the function $f''(t)$, by using the logistic second-order wavelets $\psi_2^{\alpha,\beta}$ (21)*

$$(T^{wav}f'')(c,d) = \langle f'', \psi_2^{\alpha,\beta} \rangle = \int_{-\infty}^{\infty} f''(t)\,\psi_2^{\alpha,\beta}(t)dt$$

*takes maximum (for $x_{sat} > 0$) or minimum (for $x_{sat} < 0$) value when $\alpha = a$ and $\beta = b$.*

*Proof.*

Assume that $x_{sat} > 0$. By the Cauchy-Schwartz inequality

$$|(T^{wav}f'')(c,d)| = \left|\langle f'', \psi_2^{\alpha,\beta} \rangle\right| \leq \|f''\|\|\psi_2^{\alpha,\beta}\| = \frac{x_{sat}}{\sqrt{30}a^{\frac{3}{2}}}\|\psi_2^{a,b}\|\|\psi_2^{\alpha,\beta}\| = \frac{x_{sat}}{\sqrt{30}a^{\frac{3}{2}}}$$

The maximum is reached for $\alpha = a$, $\beta = b$, because:

$$(T^{wav}f'')(c,d) = \langle f'', \psi_2^{\alpha,\beta} \rangle = \frac{x_{sat}}{\sqrt{30}a^{\frac{3}{2}}}\langle \psi_2^{a,b}, \psi_2^{a,b} \rangle = \frac{x_{sat}}{\sqrt{30}a^{\frac{3}{2}}} \tag{23}$$

Similarly we consider the case $x_{sat} < 0$.

Knowing the maximum (or minimum) value of the CWT in formula (23), we can calculate the saturation level of the corresponding logistic function (logistic wave). In the case of increasing logistic function ($x_{sat} > 0$), we get

$$x_{sat}\sqrt{30}a^{\frac{3}{2}}\max\left((T^{wav}f'')(a,b)\right) \tag{24}$$

and similarly in the case of decreasing logistic function ($x_{sat} < 0$)

$$x_{sat}\sqrt{30}a^{\frac{3}{2}}\min\big((T^{wav}f'')(a,b)\big) \qquad (25)$$

### 3.4 Algorithmic formalization

In order to make practical use of the second-order logistic wavelet $\psi_2$ (20), we implemented the family of wavelets derived from it into Matlab (Matlab Wavelet Toolbox). The complete Matlab code, including the application of these wavelets for CWT calculations, is provided in the Appendix 1.

For a given time series $(y_n)$, $n = 0,1,2,...,N + 1$ we define its central first differences

$$\Delta^1 y_n = (y_{n+1} - y_{n-1})/2, \quad n = 1,2,3,...,N$$

and the central second differences

$$\Delta^2 y_n = (y_{n+1} - 2y_n + y_{n-1}), \qquad n = 1,2,3,...,N$$

Now assume that the time series $(y_n)$ follows the logistic function $y(t) = \frac{y_{sat}}{1+exp\left(-\frac{t-b}{a}\right)}$, i.e., it has the logistic trend

$$y_n = y(n) = \frac{y_{sat}}{1 + exp\left(-\dfrac{n - b}{a}\right)}$$

and we apply Matlab's CWT to $\Delta^2 y_n$. Then for a specific range of parameters $\alpha, \beta$ the command cwt returns the value of the following sum (Index):

$$\text{Index} := \sum_{n=1}^{N} \Delta^2 y_n \psi_2^{\alpha,\beta}(n) \approx \int_{-\infty}^{\infty} y''(t)\, \psi_2^{\alpha,\beta}(t)dt$$

We can read, from the CWT scalogram, the values of parameters $\alpha, \beta$, for which the Index takes the maximum value and, based on Lemma 2, find the values of parameters $a = \alpha, b = \beta$ of the initial logistic wave $y(t)$, and, moreover, using formulae (24) or (25), we estimate its saturation level

$$y_{sat} = \sqrt{30}a^{\frac{3}{2}}\max(Index) \quad \text{or} \quad y_{sat} = \sqrt{30}a^{\frac{3}{2}}\min(Index) \qquad (26)$$

Generally we can model time series ($y_n$) by a sum of logistic functions (multilogistic function)

$$y(t) = \sum_{i=1}^{k} \frac{y_{sat}}{1+exp\left(-\frac{t-b_i}{a_i}\right)} \qquad (27)$$

$i = 1,2,...,k$, where $k$ is the number of logistic waves. If there are several overlapping logistic waves, occuring in the same time period, then the higher intensity waves (with a higher Index) may cause lower-intensity waves to be invisible on the CWT scalogram. Therefore, in order to find such waves, we can remove the first wave with the highest intensity by subtracting it from the time series ($y_n$):

$$y_n^{(1)} = y_n - \frac{y_{1,sat}}{1 + exp\left(-\frac{t-b_1}{a_1}\right)}$$

Then, for the time series $\left(y_n^{(1)}\right)$, we calculate its central second differences and for the latter we perform the CWT analysis again. The above process may be repeated several times if necessary. Usually, in order to more precisely estimate the values of some parameters of the multilogistic function, we can use some optimization methods.

### 3.5 Exact logistic wave

To illustrate the above theory, let us first consider a time series following the exact logistic trend, say with parameters $a = 10$, $b = 80$, $y_{sat} = 100,000$

$$y_n = y(n) = \frac{100{,}000}{1+exp\left(-\frac{n-80}{10}\right)}, \quad n = 0,2,3,\dots,201$$

We calculate the central second differences $\Delta^2 y_n$ (for illustration also $\Delta^1 y_n$) of the series and use for them the Matlab's CWT with the second order logistic wavelets $\psi_2{}^{\alpha,\beta}$ (21). The results are presented in Fig. 4, which shows the point (with coordinates $\beta = 80$, $\alpha = 10$) on the scalogram, with the maximum Index value (576.3). Using Lemma 2 and (26) we can estimate all parameters of the initial logistic wave:

$$a = \alpha = 10, \qquad b = \beta = 80, \qquad y_{sat} = \sqrt{30}10^{3/2} \cdot 576.3 = 99.818$$

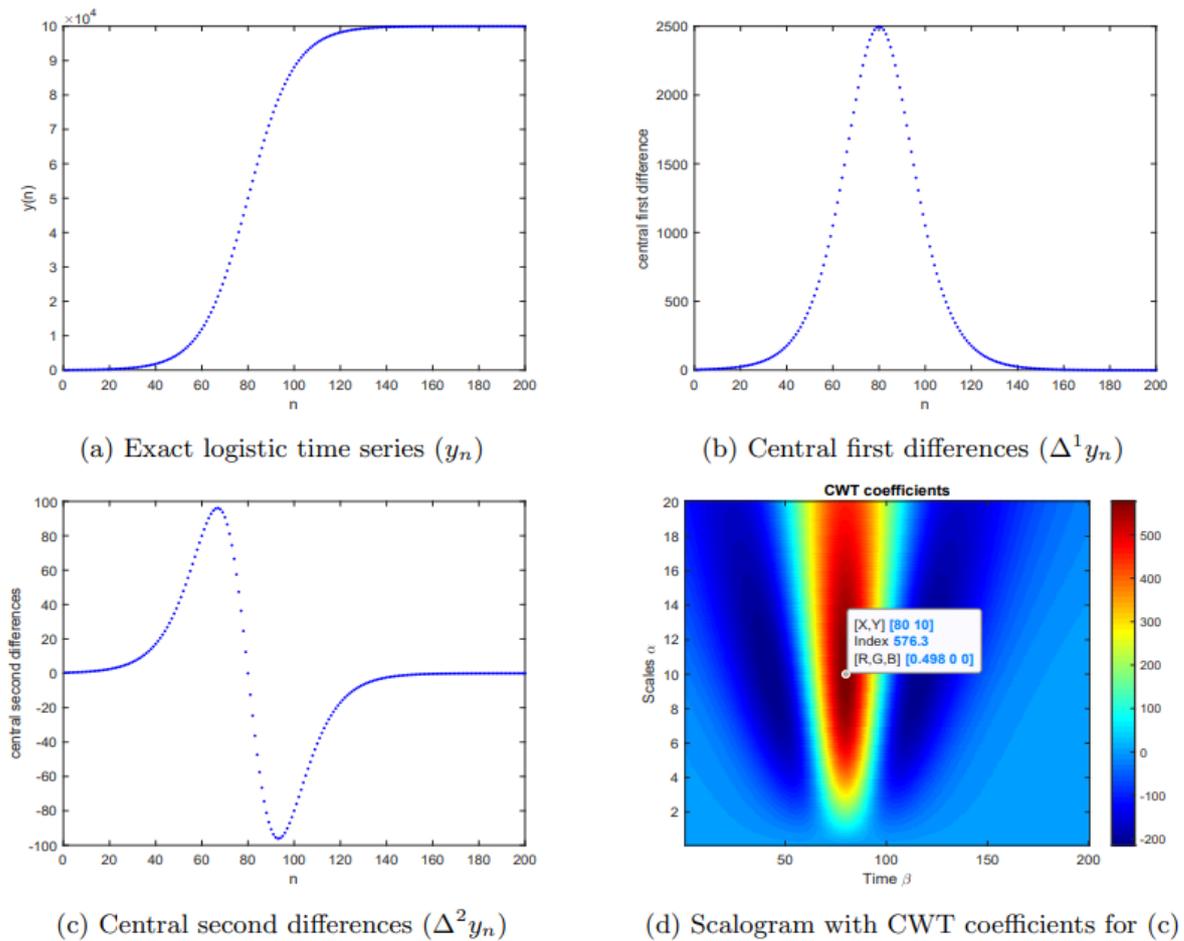

(a) Exact logistic time series ($y_n$)

(b) Central first differences ($\Delta^1 y_n$)

(c) Central second differences ($\Delta^2 y_n$)

(d) Scalogram with CWT coefficients for (c)

Figure 4: CWT analysis for the exact logistic wave

### 4. Results

In this section we analyze the data on patent applications filed between1980 and 2021 by country residents. Patent data were retrieved from the World Bank (2023). We have picked out four countries – Switzerland, USA, Germany, and Brazil. According the Global Innovation Index (GII) 2023, which counts 80 innovation indicators, Switzerland, USA and Germany rank 1st , 3rd and 8th in innovation performance above expectation for level of development in High income group of countries, Brazil is the regional innovation leader in Upper middle-income group (OECD, 2023). USA also excels in numerous GII innovation indicators.

Consider the number of patent applications filed by countries' residents between 1980 and 2021. Years are successively numbered from 1 to 42 (1980, n=1; 2021, n=42). We first calculate the central second differences and then apply the CWT to them using a family of second-order logistic wavelets $\psi_2{}^{\alpha,\beta}$ (21). Note that the range for $\beta$ is [2, 41]. The total summary data of patents is modelled by a multilogistic function (27) and a linear function (because the values of the phenomenon for both total and differential start at a positive level). The second differences, used for calculating parameters of the logistic waves, vanish for any linear function), in the following form:

$$y(t) = ct + d + \sum_{i=0}^{k} \frac{y_{i,max}}{1 + exp\left(-\frac{t-b_i}{a_i}\right)} \tag{28}$$

Differential data correspondingly are:

$$y'(t) = c + \sum_{i=0}^{k} \frac{y_{i,max}}{4a_i} cosh^{-2}\left(\frac{t-b_i}{2a_i}\right) \tag{29}$$

We will compare decomposition in Eq. (29) with redundancy predicted patterns. Redundancy distribution follows probability distribution for small $p$. Fig. 5 shows for illustrative purposes probability and redundancy distributions, where: $p = 2P\left(\frac{k}{2}\right)^2 cosh^{-2}\left(\frac{kt-\alpha}{2}\right)$; $R = -p_j log p_j$; $P = 0.05$; $k = 2$; $\alpha = 8$. Pearson correlation coefficient between two distributions is 0.988.

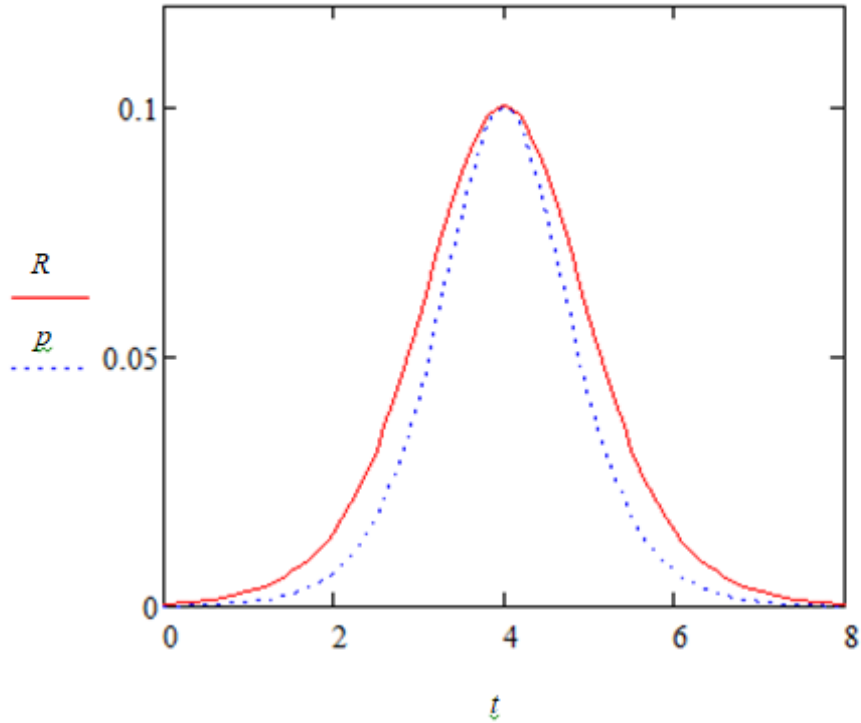

Figure 5 probability $p$ and redundancy $R$ distributions

After performing the calculations we obtained the following results for four countries: Switzerland, USA, Germany, and Brazil.

### 4.1    Switzerland, patent applications, residents

Fig. 6 shows wavelet scalograms for wave no 1 (a) and waves after removing the wave no 1 (b).

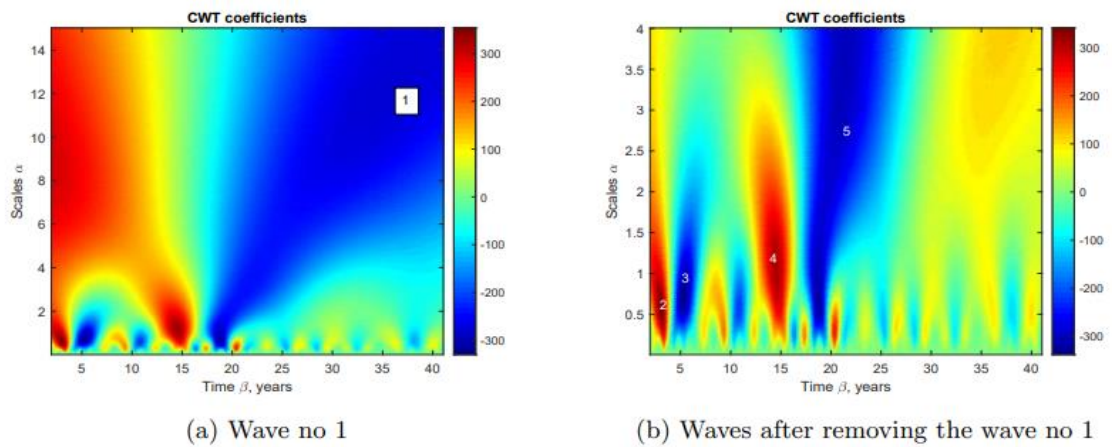

(a) Wave no 1                                          (b) Waves after removing the wave no 1

Figure 6: CWT analysis for patent applications, residents, Switzerland

Five dots correspond to five large logistic waves. We do not consider smaller waves. Calculated waves' parameters are summarized in Table 1.

| | $a$ | $b$ | $y_{sat}$ | ratio |
|---|---|---|---|---|
| wave 1 | 11 | 42 | -111018 | -60.1 |
| wave 2 | 0.5 | 3.5 | 629 | 89.9 |
| wave 3 | 0.55 | 6.4 | -546 | -38.8 |
| wave 4 | 0.89 | 15.1 | 1328 | 24.7 |
| wave 5 | 3.81 | 20.9 | -12,570 | -39.5 |

Table 1: Calculated parameters for waves 1-5 (Switzerland)

As a final result we obtain the following approximating function $y(t)$:

$$y(t) = 3059 + 3814t - \frac{111{,}018}{1 + exp\left(\frac{t - 42}{11}\right)} + \frac{629}{1 + exp\left(\frac{t - 3.5}{0.5}\right)} - \frac{546}{1 + exp\left(\frac{t - 6.4}{0.55}\right)}$$
$$+ \frac{1{,}328}{1 + exp\left(\frac{t - 15.1}{0.89}\right)} - \frac{12{,}570}{1 + exp\left(\frac{t - 20.9}{13.811}\right)}$$

After calculating $y'(t)$ we get figures Fig. 7 (a), Fig. 7 (b)

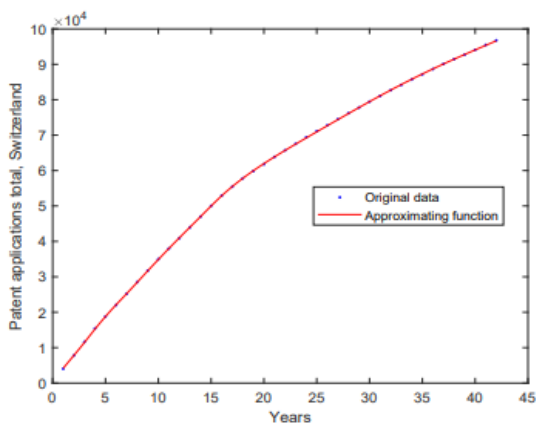

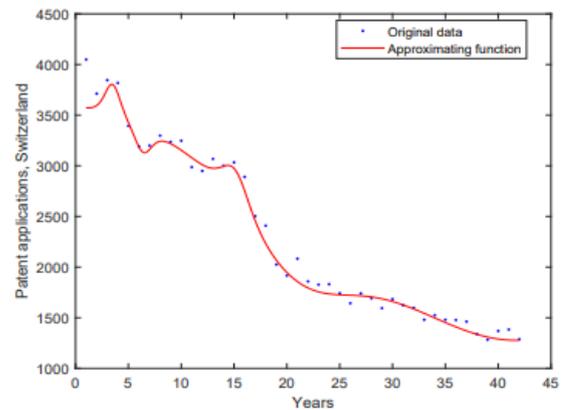

(a) Approximating function $y(t)$, $R^2 = 0.999988$   (b) Approximating function $y'(t)$, $R^2 = 0.991992$

Figure 7: Total (a) and differential (b) data for Switzerland's patents, dot line – empirical data, solid line – approximating functions

Wavelet decomposition gives one negative carrier wave (wave 1), two successive negative waves (waves 3 and 5), and two successive positive waves (waves 2 and 4) which are shown in Figs 8 - 10.

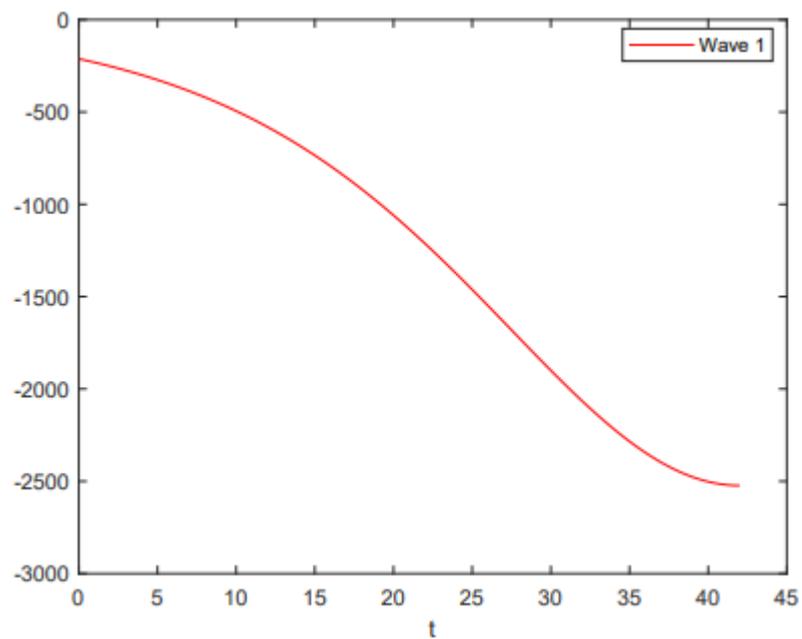

Figure 8: Wave 1; t−time

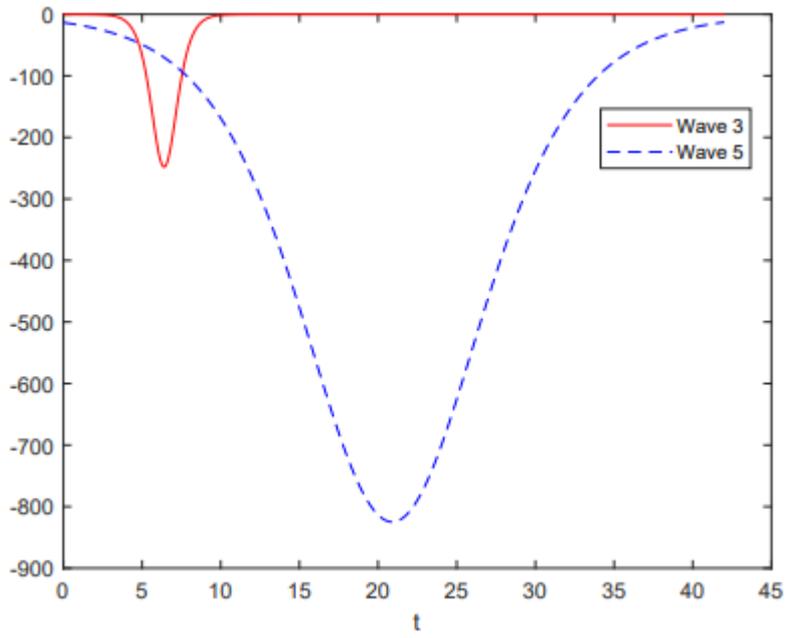

Figure 9: Waves 3 and 5; t−time

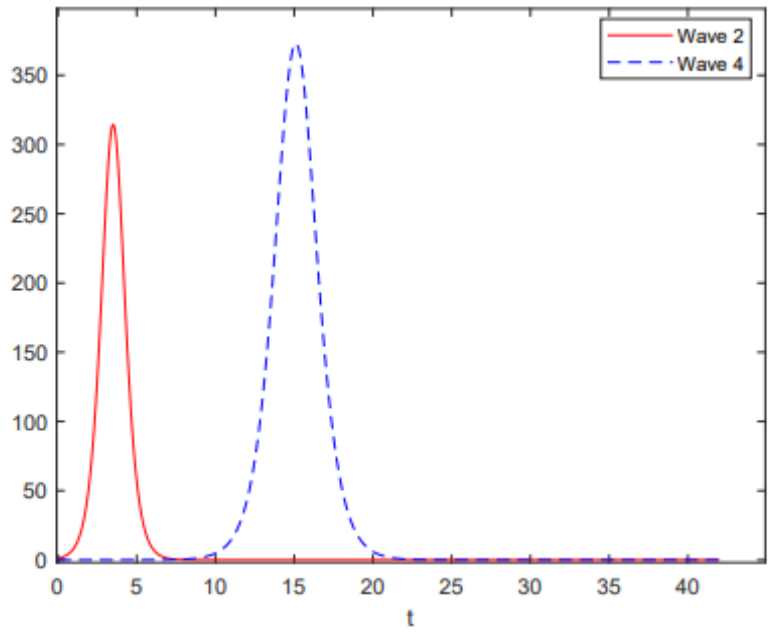

Figure 10: Waves 2 and 4; t−time

### *4.2 Germany, patent applications, residents*

Fig. 11 shows wavelet scalograms for wave no 1 (a) and waves after removing the wave no 1 (b).

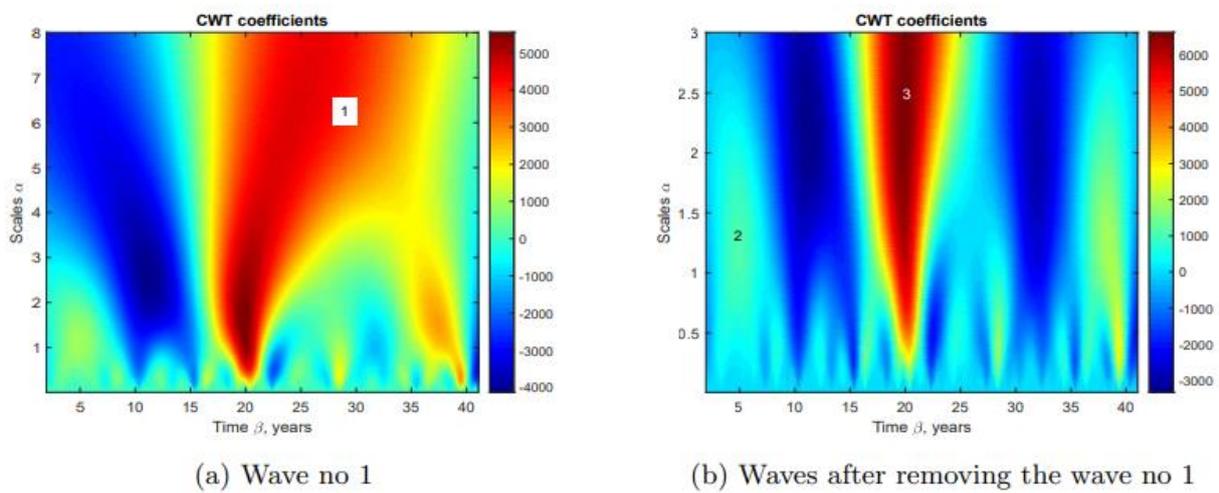

(a) Wave no 1          (b) Waves after removing the wave no 1

Figure 11: CWT analysis for patent applications, residents, Germany

Calculated waves' parameters are summarized in Table 2.

|        | $a$  | $b$  | $y_{sat}$ | ratio |
|--------|------|------|-----------|-------|
| wave 1 | 5.89 | 34.5 | 451,243   | 555   |
| wave 2 | 0.9  | 5.3  | 12,910    | 677   |
| wave 3 | 2.85 | 20.6 | 174,969   | 745   |

Table 2: Calculated parameters for waves 1-3 (Germany)

As a final result we obtain the following approximating function $y(t)$

$$y(t) = 28683t + \frac{451{,}243}{1 + exp\left(\frac{t - 34.5}{5.89}\right)} + \frac{12{,}910}{1 + exp\left(\frac{t - 5.3}{0.9}\right)} + \frac{174{,}969}{1 + exp\left(\frac{t - 20.6}{2.85}\right)}$$

After calculating $y'(t)$ we get figures Fig. 12 (a), Fig. 12 (b)

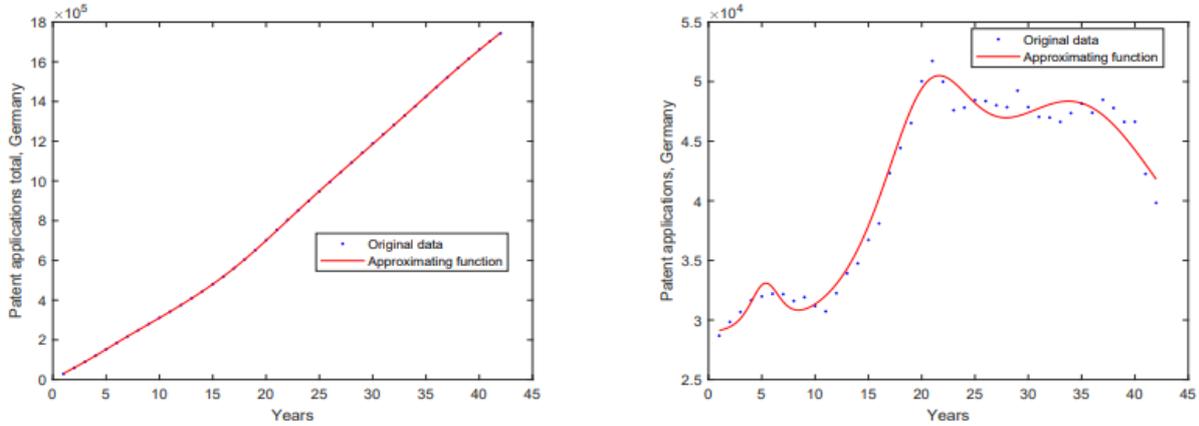

(a) Approximating function $y(t)$, $R^2 = 0.999993$     (b) Approximating function $y'(t)$, $R^2 = 0.978236$

Figure 12: Total (a) and differential (b) data for German's patents, dot line – empirical data, solid line – approximating functions

Wavelet decomposition gives one carrier linear function and three successive positive waves (1, 2, 3), which are shown in Fig 13.

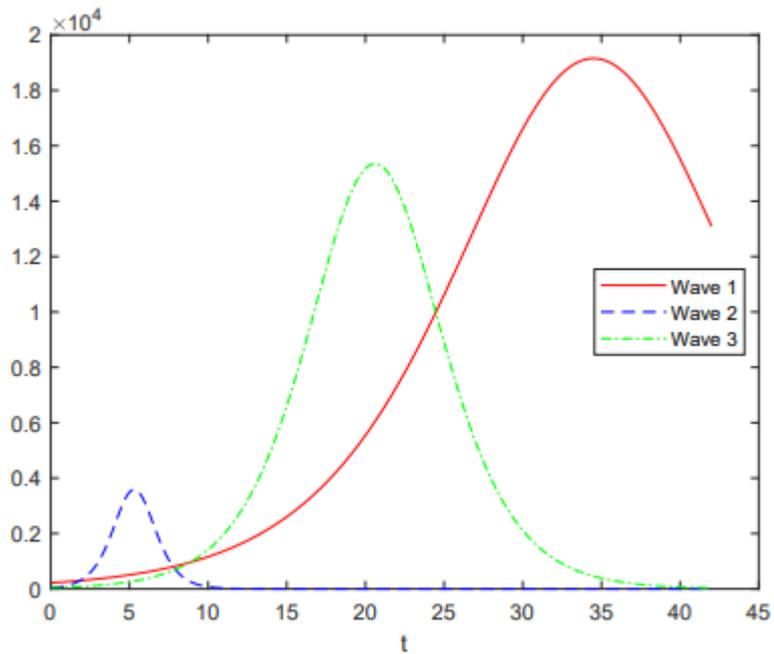

Figure 13: Waves 1, 2, and 3; t−time

### *4.2    The United States, patent applications, residents*

Fig. 14 shows wavelet scalograms for wave no 1 (a) and waves after removing the wave no 1 (b).

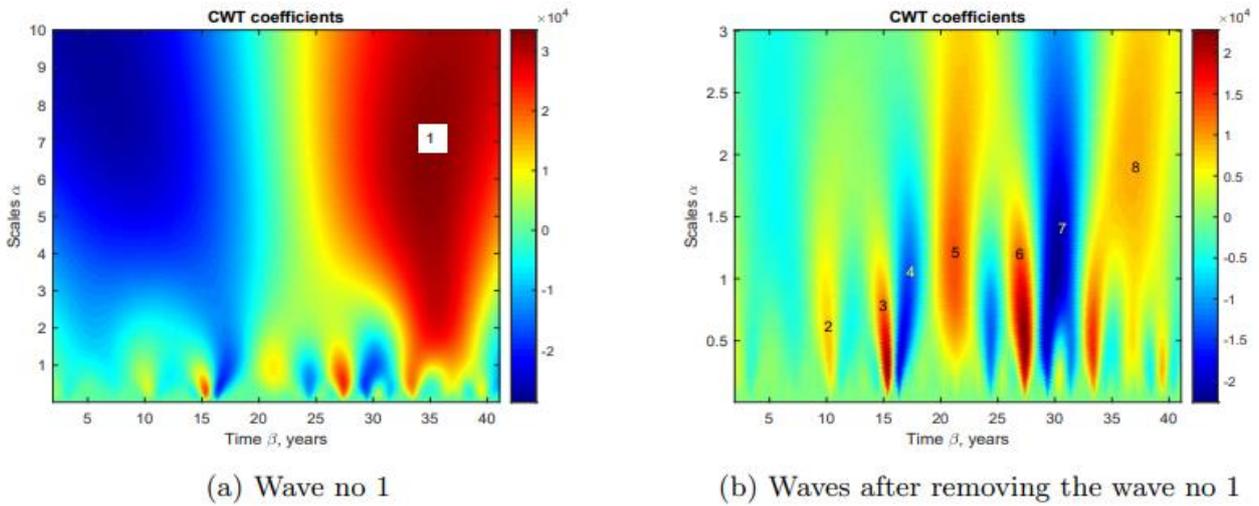

(a) Wave no 1                    (b) Waves after removing the wave no 1

Figure 14: CWT analysis for patent applications, residents, USA

Calculated waves' parameters are summarized in Table 3.

|        | $a$  | $b$  | $y_{sat}$ | ratio  |
|--------|------|------|-----------|--------|
| wave 1 | 8.07 | 35.6 | 7,404,557 | 6,443  |
| wave 2 | 0.48 | 10.3 | 14,571    | 737    |
| wave 3 | 0.38 | 15.8 | 27,328    | 1,138  |
| wave 4 | 0.58 | 16.9 | -43,911   | -1,112 |
| wave 5 | 1,16 | 21.3 | 91,217    | 923    |
| wave 6 | 0.4  | 27.4 | 31,495    | 711    |
| wave 7 | 0.95 | 31   | -113,097  | -960   |
| wave 8 | 2.9  | 38.0 | 370,000   | 839    |

Table 3: Calculated parameters for waves 1-8 (USA)

As a final result we obtain the following approximating function $y(t)$:

$$y(t) = 40{,}112t - \frac{7{,}404{,}557}{1 + exp\left(\dfrac{t - 35.6}{8.07}\right)} + \frac{14{,}571}{1 + exp\left(\dfrac{t - 10.3}{0.48}\right)} + \frac{27{,}328}{1 + exp\left(\dfrac{t - 15.8}{0.38}\right)}$$
$$- \frac{43{,}911}{1 + exp\left(\dfrac{t - 16.9}{0.58}\right)} + \frac{91{,}217}{1 + exp\left(\dfrac{t - 21.3}{1.16}\right)} + \frac{31{,}495}{1 + exp\left(\dfrac{t - 27.4}{0.4}\right)}$$
$$- \frac{113{,}097}{1 + exp\left(\dfrac{t - 31.0}{0.95}\right)} + \frac{370{,}000}{1 + exp\left(\dfrac{t - 38.0}{2.9}\right)}$$

After calculating y′ (t) we get figures Fig. 15 (a), Fig 15(b)

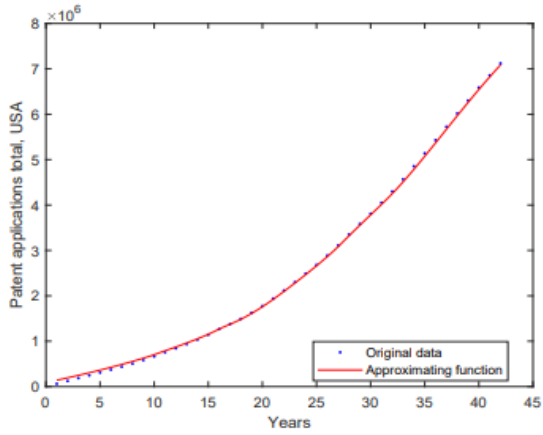

(a) Approximating function $y(t)$, $R^2 = 0.999903$

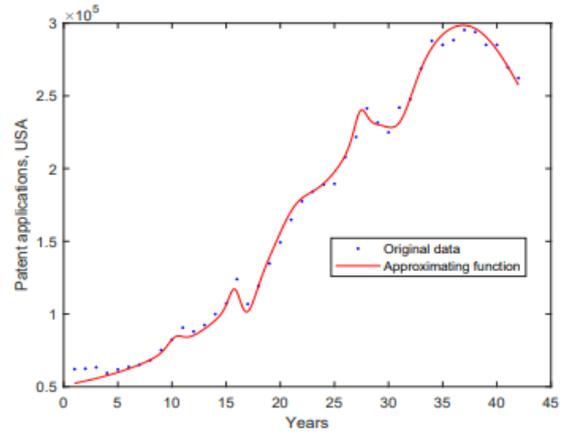

(b) Approximating function $y'(t)$, $R^2 = 0.997198$

(a)                                    (b)

Figure 15: Total (a) and differential (b) data for US patents, dot line – empirical data, solid line – approximating functions

Wavelet decomposition gives one positive carrier wave (wave 1), two successive negative waves (waves 4 and 7), and five successive positive waves (waves 2, 3, 5, 6, 8) which are shown in Figs 16-18.

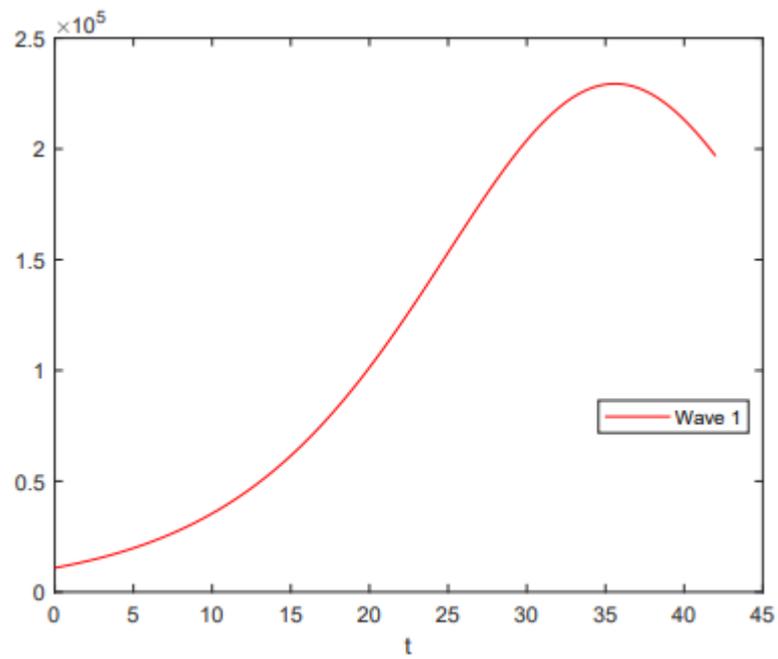

Figure 16: Wave 1; $t$ – time

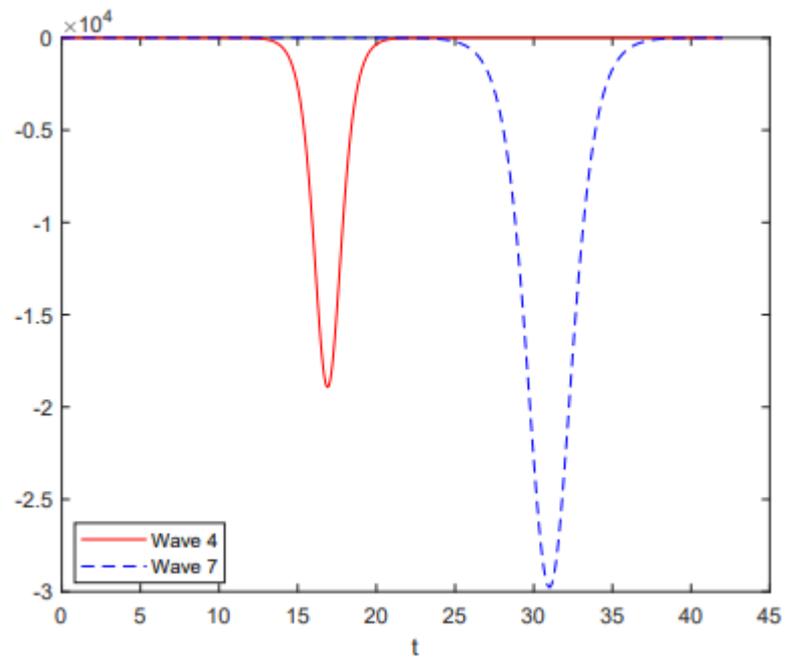

Figure 17: Waves 4 and 7; $t$ – time

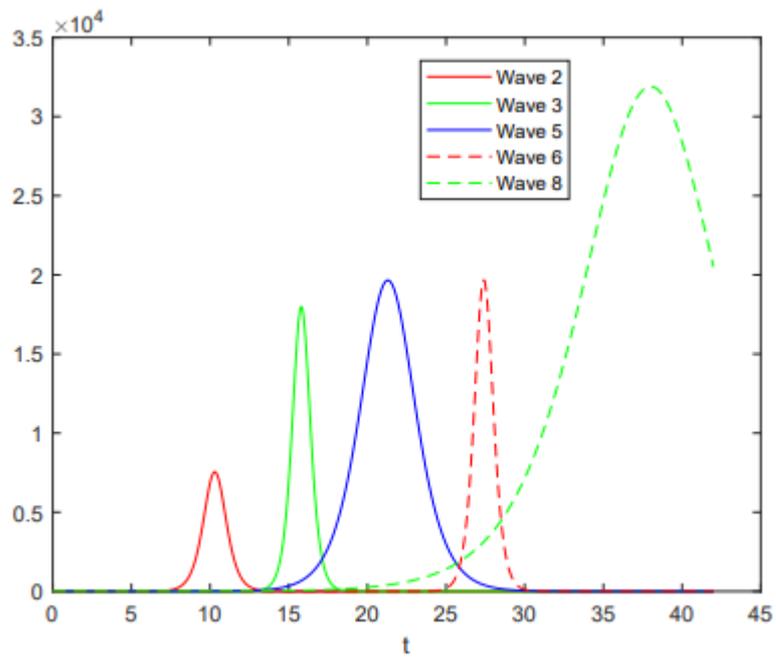

Figure 18: waves 2, 3, 5, 6, and 8; $t$ – time

### 4.4 Brazil, patent applications, residents

Fig. 19 shows wavelet scalograms for wave no 1 (a) and waves after removing the wave no 1 (b).

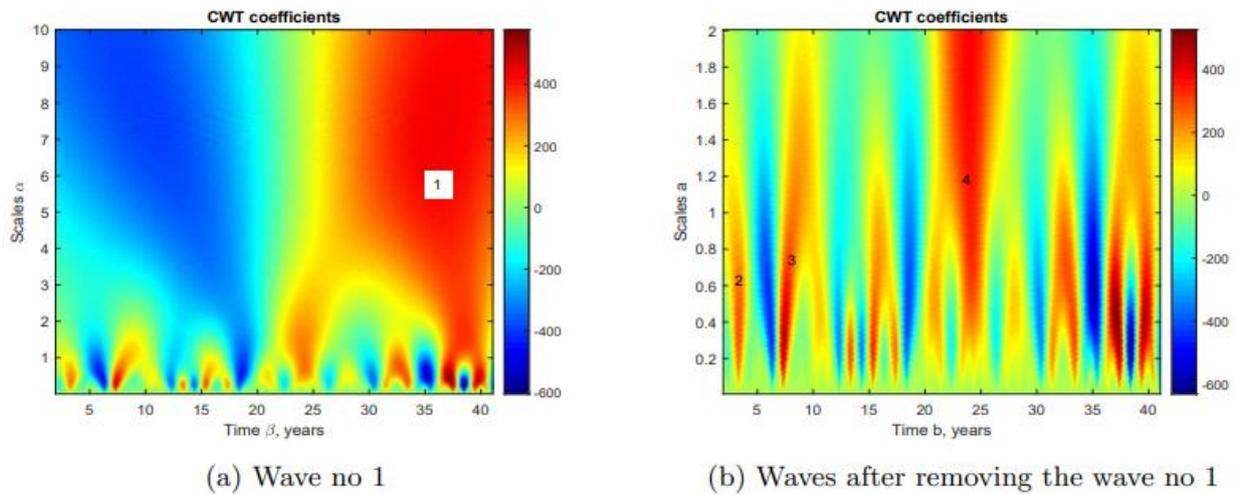

(a) Wave no 1          (b) Waves after removing the wave no 1

Figure 19: CWT analysis for patent applications, residents, Brazil

Calculated waves' parameters are summarized in Table 4.

| | $a$ | $b$ | $y_{sat}$ | ratio |
|--------|------|-----|---------|-------|
| wave 1 | 7.37 | 39 | 96,178 | 83.7 |
| wave 2 | 0.5 | 3.5 | 250 | 35.7 |
| wave 3 | 0.4 | 7.5 | 400 | 33.3 |
| wave 4 | 2 | 24 | 5000 | 26.0 |

Table 4: Calculated parameters for waves 1-4 (Brazil)

As a final result we obtain the following approximating function $y(t)$: $y(t) = 1934t -$

$$\frac{96{,}178}{1+exp\left(\frac{t-39}{7.37}\right)} + \frac{250}{1+exp\left(\frac{t-3.5}{0.5}\right)} + \frac{400}{1+exp\left(\frac{t-0.5}{0.4}\right)} + \frac{5000}{1+exp\left(\frac{t-24}{2}\right)}$$

After calculating y′ (t) we get figures Fig. 20 (a), Fig 20(b)

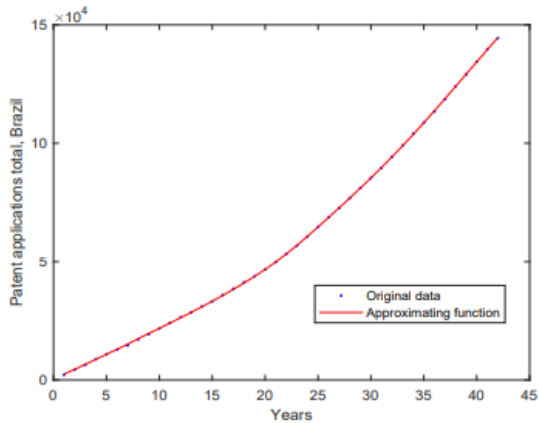 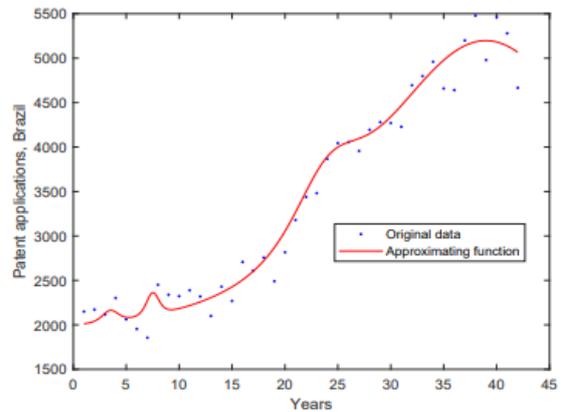

(a) Approximating function $y(t)$, $R^2 = 0.999982$    (b) Approximating function $y'(t)$, $R^2 = 0.975246$

Figure 20: Total (a) and differential (b) data for Brazil's patents, dot line – empirical data, solid line – approximating functions

Wavelet decomposition gives one positive carrier wave (wave 1), and three successive positive waves (waves 2, 3, and 4) which are shown in Figs 21, 22.

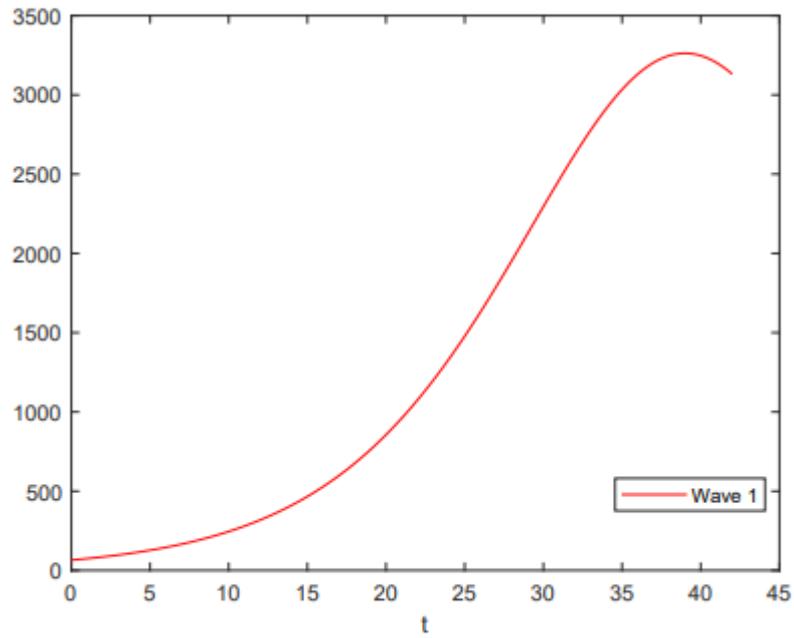

Figure 21: Wave 1; t−time

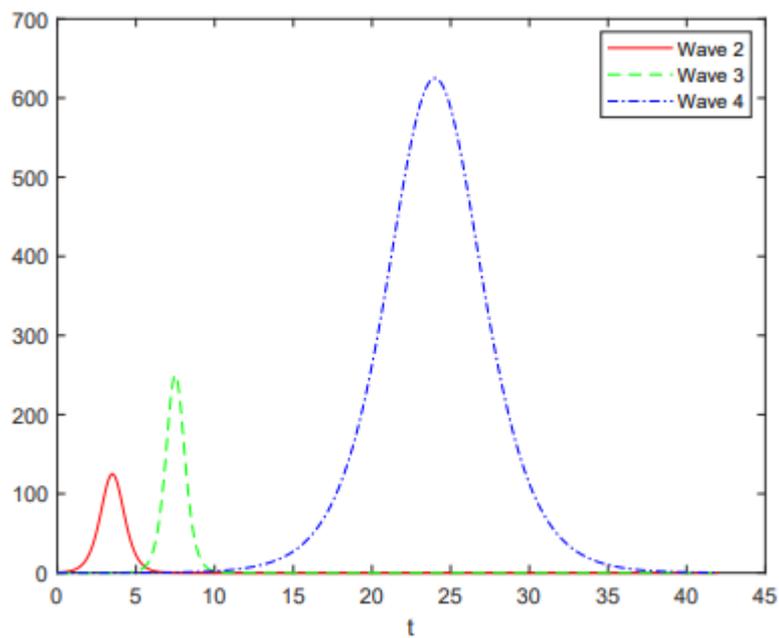

Figure 22: Waves 2, 3, and 4; t−time

Cumulative patent distributions for all the four countries follow an upward trend. Differential distributions reveal different picture. Differential dynamics can be described by one or more sequences of waves, whose tops lie on a linear trend line. One large wave can be identified, which describes a general trend, around which smaller structures formed by sequences of positive and negative waves differ.

This picture can be interpreted from the point of view of the synergy of innovation system actors' interaction. Sequences of positive waves can be attributed to positive synergy caused by regimes of self-organization occurring at the system level and leading to a decrease in the overall entropy of the system implemented at the regime level. Sequences of negative waves are caused by negative synergy leading to an increase in the overall entropy of the system as a result of historical development along the chosen trajectory (cf. Leydesdorff, 2021).

While integral trend for Switzerland is upward differential trend is downward indicating weakening of patent activity. This weakening may indicate imbalance in NIS performance. CWT analysis allows select five waves. The biggest one (wave 1) is a carrier wave. Two competing dynamics can be distinguished. Negative waves (3 and 5) may be related to two different historical trajectories. Positive waves (2 and 4) indicate synergy generated at the regime level.

Germany demonstrates positive dynamics at both integral and differential scales. We cannot identify wave sequences, which means that the dynamic is different from that described by Eq. 6.

USA case is the most representative. Both positive (waves 2, 3, 5, 6, 8) and negative (waves 4 and 7) dynamics can be identified. We can identify three regimes. Waves 4 and 7 form a negative sequence. Wave peaks coincide with the dates of the Asian financial crisis of 1998, as well as the LTCM hedge fund crisis in the United States in 1998, and the crisis of 2008. There is an opinion that the subsequent crisis may be a logical continuation of the previous one. Here we see a picture when two successive waves form a chain described by the solution of one equation. Two positive trends can also be identified. First formed by waves 2, 5, 8, and second formed by waves 3 and 6. The peak of wave 8 corresponds 2020 year (the start of Covid 19 pandemy). This indicates that patent issuance is currently in decline. It is interesting to mention that development of waves 3 and 6 coincides with the development of Rao-Stirling diversity in IPC, calculated with respect to patents related to nine material technologies for photovoltaic cells which were registered by USPTO (Leydesdorff *et al.*, 2015). Wave sequences suggest that technology develops in cycles. Standard lifecycle includes early stage, development phase, and commercial phase.

The case of Brazil is structurally similar to the example of Germany. The peaks of three successive waves form a linear trend. The last observed peak corresponds 2020 year. Since then there has been a decline in the level of synergyю Absence of synergy means absence of innovations and correspondingly absence of patents.

## 5. Conclusion

In this paper, we present a procedure for decomposing the dynamic evolution of patents through the logistic Continuous Wavelet Transformation. We analyze the aggregate longitudinal distributions of patents in terms of logistic functions. The implemented method allows us identify the cyclic structure of differential distributions, which are presented as a sum of derivatives of logistic functions that coincide with soliton solutions of the KdV equation. Another contribution of the paper is that we provide a theoretical basis for the technique used. The KdV equation appears naturally in the TH model of innovation, which analyzes the evolution of an innovation system in terms of three sub-dynamics. The information contained in the configuration of the three sub-dynamics is an indicator of the synergy between actors. Given that patents are an outcome of the innovation system, one can assume that patenting activity is proportional to the effectiveness of the innovation system. We analyze patent dynamics in terms of a measure of synergy as an expected knowledge base in an economy.

We zoom in on four countries – Switzerland, USA, Germany, and Brazil. By comparing different countries we can explore different dynamics of corresponding innovation systems, provided that patent production is proportional to the strength of innovation system. According to our findings patent trends follow solitary wave patterns, comprised of positive and negative chains of solitons which represent fundamental dynamics of patent wave in terms of innovation system systemness. Systemness is estimated as synergy, which is decisive for the strength of innovation system (Fritsch, Slavtchev, 2006). With respect to system synergy two competing processes are present – historical generation of variations and interaction among sub-dynamics as selection environments. System uncertainty is generated in historical process of variation. Interaction among selection environments leads to generation of new options, so that total number of options is increased. Additional redundancy leads to the reduction of uncertainty (Petersen, Rotolo, & Leydesdorff, 2016). The trade-off between these processes defines the synergy sign. Positive synergy indicates that generation of Shannon-type information prevails and system uncertainty is increased. Negative synergy implies that non-linear generation of redundancy in loops of communication comes out on top, which means the reduction of uncertainty. Innovation system studies show that synergy can be generated at various structural levels, such as national or regional, i.e. innovation system simultaneously operates in one or two regimes (e.g. Leydesdorff, et al., 2015; Gomez, Iturriagagoitia, & Leydesdorff, 2019; Almeida, Porto-Gómez, & Leydesdorff, 2023). Our results confirm the finding that NIS can simultaneously function in different regimes, which are represented by different trends and moreover the number of trends can be more than two.

The approach we use allows us assess the quality of national innovation systems in terms of synergy, based on entropy statistics. Absence of synergy means absence of innovations and correspondingly absence of patents. It also proves possible to predict the propagation and development of patent waves, especially when the wave changes from expansion to decline. This prediction may have significant economic and technological consequences.

Combination of logistic continuous CWT transform and TH metaphor allows one to conceptualize and implement the approaches, used in non-linear dynamics, such as KdV equation and its mathematically similar equations, such as e.g. non-dissipative Lorenz model (1963, a and b), to study the dynamics of the NIS with respect to its predictability and stability analysis.

**Acknowledgement**

Inga Ivanova acknowledges support from the Basic Research Program at the National Research University Higher School of Economics.

**Funding statement**

The research of the author G.R. was partially funded by the 'IDUB against COVID-19' project granted by the Warsaw University of Technology (Warsaw, Poland) under the program Excellence Initiative: Research University (IDUB), grant no 1820/54/201/2020.

**Conflict of Interests**

The authors declare that there is no any conflict of interest in the submitted manuscript.

**Appendix 1**

Matlab Program codes

Listing 1: The logistic mother wavelet $\psi_2$ function, M-file logist.m:

```
function [psi,t] = logist(LB,UB,N,~)
%LOGISTIC Logistic wavelet.
% [PSI,T] = LOGIST(LB,UB,N) returns values of
% the Logistic wavelet on an N point regular
% grid in the interval [LB,UB].
% Output arguments are the wavelet function PSI
% computed on the grid T.
% This wavelet has [-7 7] as effective support.
% See also WAVEINFO.
% Compute values of the Logistic wavelet.
t = linspace(LB,UB,N); % wavelet support.
psi =sqrt(30)* (exp(-2*t)-exp(-t))./(1+exp(-t)).^3;
end
```

Listing 2: Code for adding the logistic wavelet family (with the family short name fsn 'lgs')

```
>>wavemngr('add','Logistic','lgs',4,'','logist',[-7 7]);
```

Listing 3: Code for calculating CWT coefficients and producing scalogram (length of the time series N)

```
>>CWTcoeffs = cwt(FileName,1:10,'lgs');
t = linspace(1,N,10000);
imagesc(t,1:10,CWTcoeffs);
colormap jet;
axis xy;
colorbar;
title('Title');
xlabel('Time \beta, years');
ylabel('Scales \alpha');
```

**Appendix 2**

Equation (6):

$$P_T + 6PP_X + P_{XXX} + C_1 = 0$$

after transition to a moving frame: $t = X - T$ and integration is written as follows:

$$P_{tt} + 3P^2 - P + C_1 t = 0 \qquad (A2.1)$$

here $P$ stands for probability density. Corresponding redundancy (here we do not define a sign, which can be either positive or negative): $R = PlnP$, by setting $P = e^q$ ($q < 0$), can be written in the form: $R = qe^q$. This expression can be inversed: $q = W(R)$, where $W(R)$ is the Lambert function (e.g. Lehtonen, 2016). Accordingly:

$$P = e^W = \frac{R}{W} \qquad (A2.2)$$

Differentiating (A2.2) with respect to *t* and taking into account that:

$$W_t = \frac{W}{R(1+W)} R_t = \frac{R_t}{R + e^W}$$

$$W_{tt} = \frac{R_{tt}(R + e^W) - R_t{}^2 - R_t W_t e^W}{(R + e^W)^2}$$

we after some transformations get the expression:

$$P_{tt} = e^W W_t^2 + e^W W_{tt} = \frac{WRR_{tt}\left(1+\frac{1}{W}\right) - W\left(\frac{1}{1+W}\right)R_t^2}{R(1+W)^2}$$

Substituting the resulting expressions into equation (A2.1) we get:

$$\frac{W^3 RR_{tt}\left(\frac{1+W}{W}\right) - W^3 R_t^2 + 3R^3(1+W)^2 - R^3(1+W)^2}{W^2 R(1+W)^2} + C_1 t = 0 \qquad (A2.3)$$

in a linear approximation for small $R$ values $W(R) \sim R$ Eq. (A2.3) becomes:

$$\frac{R_{tt} - \left(\frac{R_t}{1+R}\right)^2 + 2(1+R)}{(1+R)} + C_1 t = 0 \qquad (A2.4)$$

The second term in the numerator is the square of the derivative of $lnR$: $(lnR)_t = \frac{R_t}{1+R}$. Expanding the logarithm $R$ in a Tailor series, taking the derivative, squaring and preserving terms up to the second order of smallness, we obtain:

$$(lnR)_t^2 \sim 1 - 2R + R^2 + \cdots \qquad (A2.5)$$

Substituting (A2.5) into (A2.4), we get:

$$R_{tt} - R^2 + 4R + 1 + C_1 t = 0 \qquad (A2.6)$$

Eq. (A2.6) can be derived from the equation:

$$4R_T - 2RR_X + R_{XXX} + C_1 = 0 \tag{A2.7}$$

upon differentiation and transition to a moving frame $t = X + T$.